# Mathematical analysis of a mouse experiment suggests little role for resource depletion in controlling influenza infection within host

Hasan Ahmed[1], James Moore[1], Balaji Manicassamy[2], Adolfo Garcia-Sastre[3], Andreas Handel[4], Rustom Antia[1]*

[1]Emory University, Atlanta, Georgia, USA
[2]University of Chicago, Chicago, Illinois, USA
[3]Icahn School of Medicine at Mount Sinai, New York, New York, USA
[4]University of Georgia, Athens, Georgia, USA
*Corresponding author, rantia@emory.edu

## Abstract

How important is resource depletion (e.g. depletion of target cells) in controlling infection within a host? And how can we distinguish between resource depletion and other mechanisms that may contribute to decline of pathogen load or lead to pathogen clearance? In this paper we examine data from a previously published experiment. In this experiment, mice were infected with influenza virus carrying a green fluorescent protein reporter gene, and the proportion of lung epithelial cells that were influenza infected was measured as a function of time. Three inoculum dose groups - $10^4$ PFU, $10^6$ PFU and $10^7$ PFU - were used. The proportion of cells infected was estimated to be about 21 (95% confidence interval: 14-32) fold higher in the highest dose group than in the lowest dose group with the middle dose group in between. We show that this pattern is highly inconsistent with a model where target cell depletion is the principal means of controlling infection, and we argue that such a pattern constitutes a reasonable criterion for rejecting many resource depletion models. A model with an innate interferon response that renders susceptible cells resistant fits the data reasonably well. This model suggests that target cell depletion is only a minor factor in controlling natural influenza infection.

## Introduction

It is well known that adaptive immunity (antibodies and T cells) can control influenza and other infections within host [1]. Yet experimental data, such as the data analyzed in this paper [2] as well as other data [3] [4], shows that primary infection often peaks very early - within a few days of infection - even in naive hosts. In these instances the infection begins to decline substantially earlier than what might be expected if adaptive immunity were the main factor in controlling infection [5] [6]. This suggests that innate immunity may be responsible for controlling infection with adaptive immunity becoming important some days later and helping to clear the infection. An alternative explanation is resource depletion. In this framework the infection uses up some resource - for example target cells in the case of viral infection - and depletion of this resource is what controls infection.

In fact target cell depletion has been proposed as a mechanism for controlling influenza infection [7]. The model has been shown to fit influenza virus load dynamics well [7]. And mathematical modeling studies previously showed that the viral load kinetics can in general be well described by the target cell depletion mechanism (reviewed in e.g. [8] [9]). But the model may seem scientifically unsatisfying in that it suggests that the adaptive and innate immune responses have a very limited role in controlling infection. Either the infection does not take off ($R_0$ less than 1) or a majority of susceptible cells are killed ($R_0$ substantially greater than 1) - only for a narrow range of $R_0$ is other behavior observed; here $R_0$ is the basic reproduction number for cell to cell transmission within a host [10].

Saenz et al [3] have analyzed data of influenza infection in ponies. They observed cell death of only 27% at the end of influenza infection and argue that target cell depletion cannot explain this result. But latent heterogeneity in which only a subset of supposedly susceptible cells are actually susceptible to infection [11] could easily explain this result. In fact the data from Saenz et al can be fit almost perfectly with a target cell depletion model under the assumption that only a fraction of lung epithelial cells are actually susceptible (see supplement S1). Therefore another means of distinguishing between target cell depletion and other models is needed.

One way of distinguishing between target cell depletion and immunity in controlling infection is by perturbing (typically depleting) various components of the immune system. Comparing different mechanistic models with data from such experiments has provided one way of determining the role of different components of immunity in the control of influenza virus infection [6]. However, experiments which deplete components of the immune system are not ethically feasible in humans. In contrast human infection challenge studies that vary the inoculum dose may be more ethically feasible.

In this paper we build on earlier models that explore the impact of innoculum dose on the dynamics of infection [12]. We show that the relationship between inoculum dose and final size (the total number of cells infected during infection) can allow rejection of the resource depletion model. We then show that based on this criterion data from a mouse influenza experiment strongly suggests mechanisms other than resource depletion. Finally we describe an interferon model that fits this data well. This interferon model suggests that for lower more physiological inoculum doses target cell depletion plays only a miniscule role in limiting influenza infection.

# Results and discussion

## Criterion for distinguishing between resource depletion and other models

One of the simplest resource depletion models is the classical susceptible infected removed (SIR) model developed by Kermack and McKendrick [13]. This model was originally proposed to model between host spread of an infection within a community. A mathematically equivalent and conceptually analogous model can be used to model between cell spread of infection within a host.

$$dS/dt = -r * S * I$$

$$dI/dt = r * S * I - b * I$$

$$dR/dt = b * I$$

Here S is the number of susceptible cells. I is the number of infected cells, and R is the number of removed (i.e. dead or otherwise unavailable) cells. (For between host models R some times stands for recovered.) Since R=N-S-I it is not necessary to explicitly model all three compartments; here N is the number of susceptible plus infected cells at time zero i.e. at the beginning of the infection. (This model does not include an explicit equation for the parasite; infected cells produce new infected cells via the unmodeled production of parasites.)

The final size (F) is the total number of cells infected during the infection. For the Kermack Mckendrick model the final size is given by

$$F = N * (1 + W(-R_0 * (1 - I(0)/N) * e^{-R_0})/R_0)$$

where W is the upper branch of the Lambert W function and I(0) is the number of infected cells at time zero. $R_0$=N*r/b. (See supplement S2 for a derivation of this final size formula.) In this model I(0) corresponds to inoculum dose.

The maximum fold change (M) in final size due to changes in I(0) can be derived from the final size formula.

$$M = (1 + W(-R_0 * e^{-R_0})/R_0)^{-1}$$

Inverting this equation gives

$$R_0 = -ln(1 - 1/M) * M$$

. As shown in figure 1 M decreases as $R_0$ increases such that M approaches 1 for higher values of $R_0$.

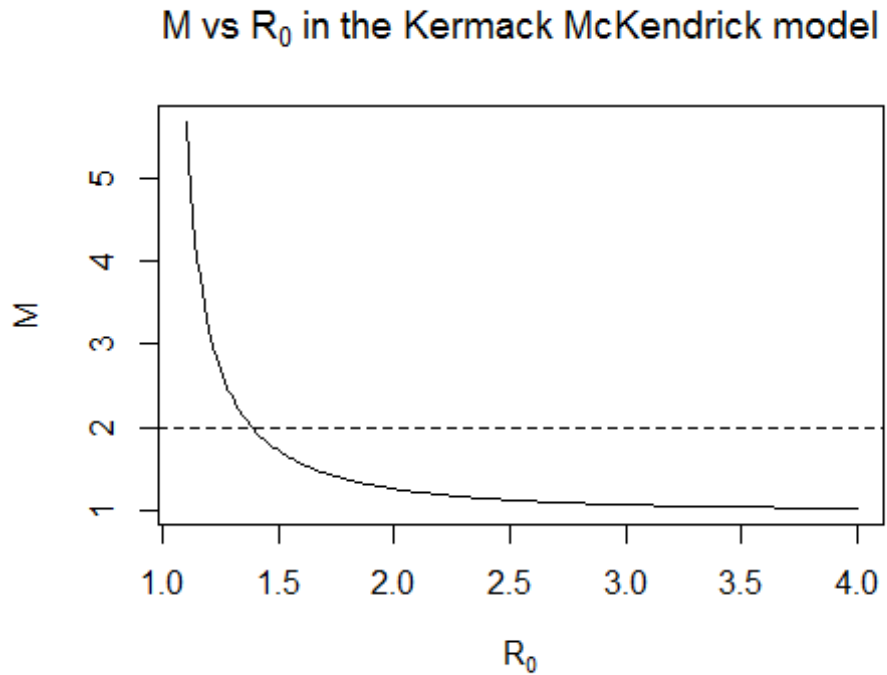

Fig 1. M vs $R_0$ in the Kermack McKendrick model. M, the maximum fold change in final size due to changes in inoculum dose, is close to 1 in the Kermack McKendrick model except at very low values of $R_0$. Hence M>2 seems a reasonable criterion for rejecting models like the Kermack McKendrick model where resource depletion is the only means of control.

Remarkably the Kermack McKendrick final size formula also gives the final size for more complicated resource depletion models; this includes models in which infected cells progress through multiple infectious stages and those in which certain cell populations are more infectious than others [14] [15]. Moreover explicitly including a parasite compartment requires only a small change to the final size formula (to accommodate an initial condition than includes parasite count) whereas the formula for M remains unchanged (supplement S3).

Therefore the relationship between inoculum dose, $R_0$ and F is a means of distinguishing between target cell depletion and other mechanisms of control. In the variety of models mentioned in the previous paragraph M>2 requires $R_0$<1.39. Such low values of $R_0$ are unlikely for several reasons. 1) These values overlap only a small portion of the viable range of $R_0$; $R_0 \leq 1$ being inviable. 2) Such low $R_0$ suggests lack of robustness; conditions that are even slightly unfavorable to the parasite could cause $R_0$ to drop to 1 or below. 3)

Low $R_0$ corresponds to high risk of early stochastic extinction (i.e. stochastic extinction when only a small number of cells have been infected); under typical assumptions the probability of early stochastic extinction is $1/R_0$ [16]. Therefore we suggest that greater than 2 fold increase in final size from increase in inoculum dose is a reasonable criterion to suggest that mechanisms other than resource depletion play an important role.

## Target cell depletion cannot explain Manicassamy data

Here we examine data from Manicassamy et al [2] augmented with previously unpublished data from the same experiment. In this experiment, naive mice were infected with influenza virus carrying a green fluorescent protein gene, and the proportion of lung epithelial cells that were influenza infected was measured. Three inoculum dose groups - $10^4$ PFU, $10^6$ PFU and $10^7$ PFU - were used. As expected from the study design I(0) seems to increase with dose (figure 2).

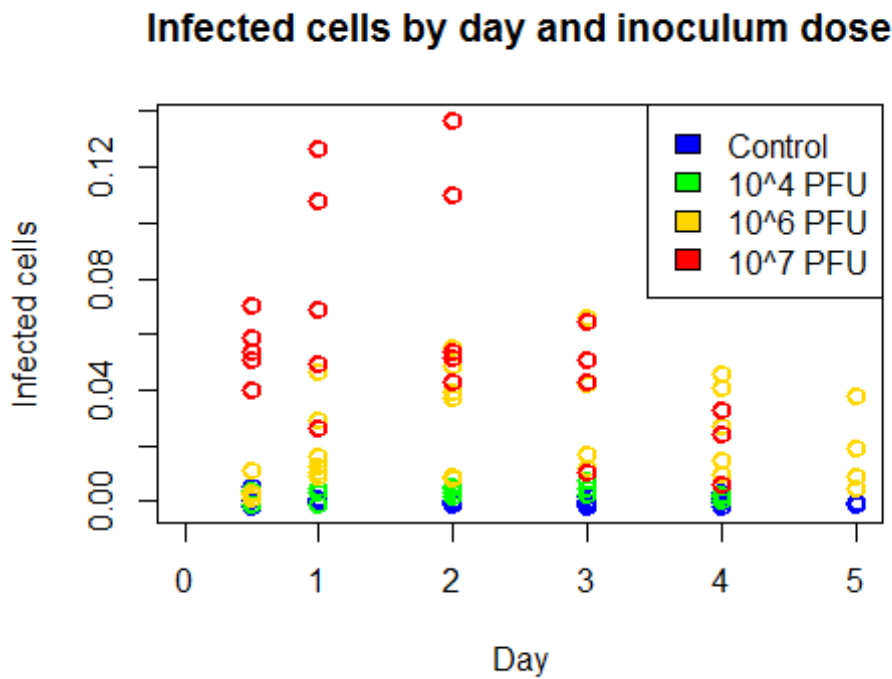

**Fig 2. Infected cells by day and inoculum dose.** This figure shows the proportion of lung epithelial cells that are influenza infected according to green fluorescent protein expression. Proportions were adjusted for background fluorescence by subtracting the mean of the control group. Infected proportions appear to increase with dose.

The final size is proportional to the integral or area under the curve (AUC) of infected cells over time. Because the infection seems to be clearing by day 4 and because for the $10^4$ PFU and $10^7$ PFU groups we only have data until day 4, we use the AUC over the first 4.5 days to approximate the final size.

The AUC is much higher in the $10^7$ PFU group as compared to the $10^4$ PFU group (~21 fold difference, 95% confidence interval: 14-32). The AUC in the $10^6$ PFU group is also much higher than in the $10^4$ PFU group (~9.1 fold difference, 95% confidence interval: 6.1-14). The difference in AUC between the two higher dose groups is also statistically significant (~2.3 fold difference, 95% confidence interval: 1.7-3.2). Figure 3 shows the AUCs by dose group.

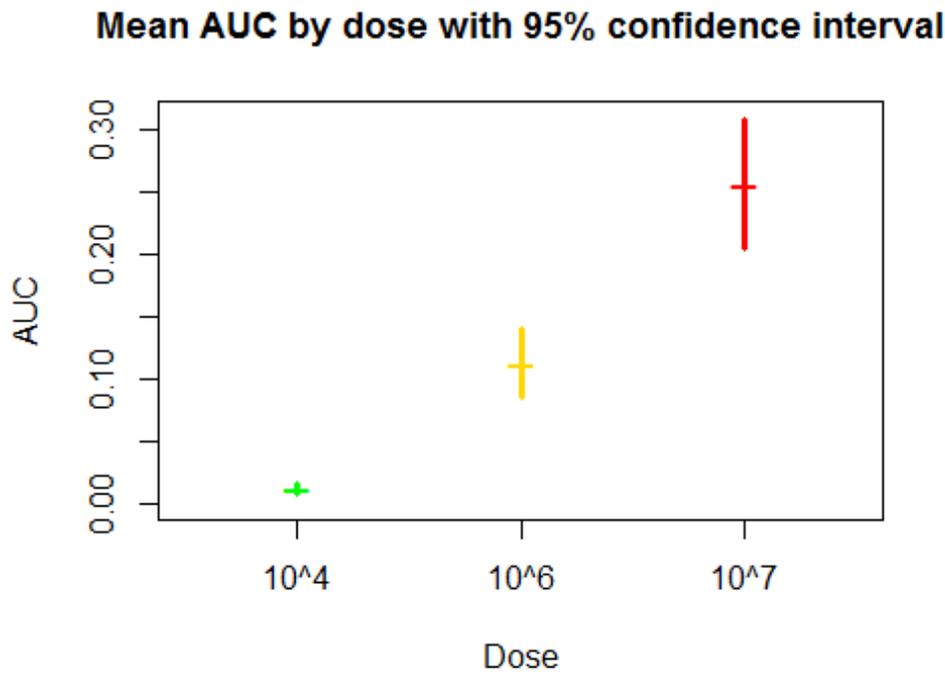

Fig 3. Mean AUC by dose with 95% confidence interval. AUC (infected cells over time) is much higher in the high and middle dose groups compared to the low dose group. The difference between the two higher dose groups is also statistically significant (p<0.05). This pattern, large variation in AUC with inoculum dose, is suggestive of control mechanisms other than resource depletion.

Therefore this data is highly inconsistent with the Kermack McKendrick model unless $R_0$ is implausibly small ($R_0<1.04$). In contrast Baccam et al [7] estimate an $R_0$ of 22 for influenza. Furthermore since the initial growth rate, $d\ln(I)/dt$ at time zero, equals $((1-I(0)/N)*R_0-1)*b$ a very high death rate (b) is needed to fit the early growth rate if $R_0$ is low.

## Model with interferon response can explain data

Because the infection peaks around or before day 3 in all three inoculum dose groups and because this is primary infection data, it seems unlikely that adaptive immunity is a major factor in controlling the infection. Because influenza is a viral infection, the role of phagocyte activation seems relatively less important. Therefore we propose a model in which type I interferon causes susceptible cells to become resistant to infection similar to [3].

$$dS/dt = -r * S * I - X$$

$$dI/dt = r * S * I - b * I$$

$$dR_1/dt = b * I$$

$$dR_2/dt = X$$

Here $R_1$ is dead cells. $R_2$ is cells that have become resistant in response to interferon. X is interferon mediated conversion of susceptible cells to resistant. If X=k*S*I then the model is mathematically equivalent to a Kermack McKendrick model and therefore suffers from the same limitations. Realistically, infected cells produce interferon with some delay relative to time of infection, interferon diffuses and after some time, depending on the local concentration of interferon, susceptible cells convert to a resistant phenotype. Therefore accurately modeling X may involve multiple delays and spatial consideration. However, for this data $X = k * S * e^{-b*l_1} * I(t - l_1 - l_2)$ fits reasonably well. Here infected cell produce interferon with lag $l_1$ and susceptible cells respond to interferon with lag $l_2$. Because of lack of identifiability we assume $l_1=l_2$ so the model simplifies to $X = k * S * e^{-b*l/2} * I(t - l)$ where $l=l_1+l_2$.

For fitting the model we used the mean for each group at each time point. As shown in figure 4 the model fits the means well; table 1 shows the parameter values used.

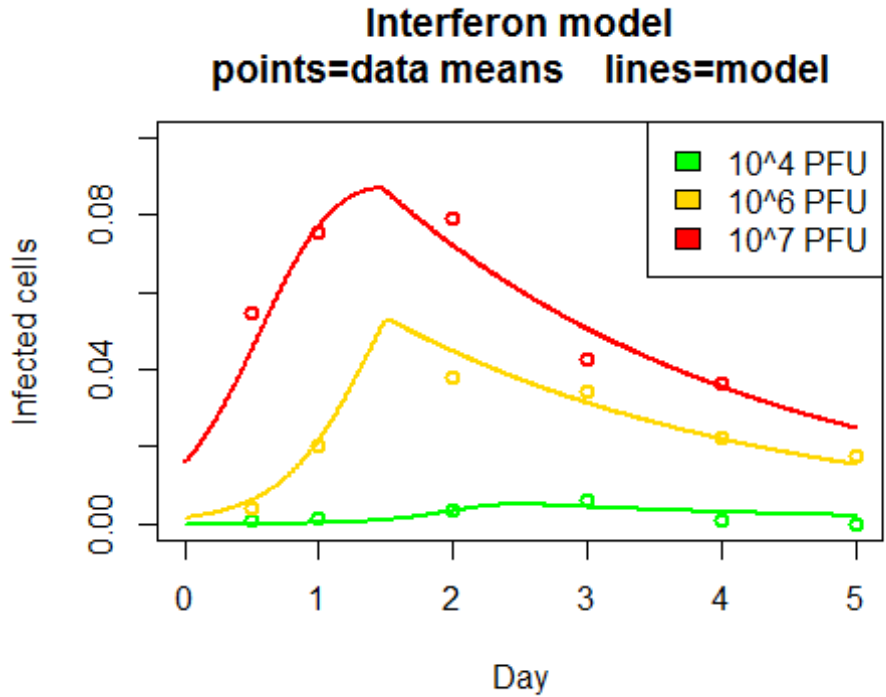

Fig 4. Interferon model. The interferon model closely matches the means of each group.

Table 1. Parameter values for interferon model

| Parameter | Value | Units |
| --- | --- | --- |
| I(0) | 0.000016, 0.0016, 0.016* | proportion of lung epithelial cells |
| N | 0.132 | proportion of lung epithelial cells |
| r | 24.1 | 1/days |
| b | 0.355 | 1/days |
| k | 28500 | 1/days |
| $l_1+l_2$ | 1.47 | days |

All values are based on fitting to the data. See methods section for more information. *For low, middle and high dose groups respectively; 1:100:1000 ratio was forced.

### Implications of interferon model

Because all three doses in the Manicassamy experiment were lethal, we use half of the lowest inoculum dose for this section (i.e. we use I(0)=0.000008). For consistency with earlier sections and because adaptive immunity is likely to clear the infection, we focus on the first 4.5 days. We assume that the severity and transmissibility of infection is given by

AUC over the first 4.5 days. We assume that k and $l_2$ are purely host dependent parameters, so we focus on N, r, b and $l_1$ which we consider to be dependent on both virus and host. Table 2 shows the impact of these parameters on AUC.

**Table 2. Influence of model parameters**

| Parameter | dln(AUC)/dln(parameter) |
|---|---|
| N | 5.42 |
| r | 5.36 |
| b | -0.72 |
| $l_1$ | 1.97 |

The model suggests that N and r have the greatest impact on transmissibility and severity and $l_1$ also has a large impact. These results are consistent with claims that the highly pathogenic 1918 H1N1 strain more readily infected cells deep in the lung (higher N), replicates to higher titers in vitro (higher N or r) and more effectively blocked interferon production (higher $l_1$) [17] [18]. b has a smaller but still substantial impact. This is consistent with the cytolytic nature of influenza. Since a large drop in b would be offset by a small drop in N, r or $l_1$, evolution would favor a cytolytic strain with slightly higher N, r or $l_1$ over a noncytolytic strain.

In this scenario the role of target cell depletion is miniscule; depletion of target cells reduces AUC by 5% compared to an otherwise equivalent model in which infection does not reduce the number of susceptible cells. In contrast interferon reduces AUC by 92%. This suggests that resource depletion plays only a minor role in controlling influenza infection.

# Concluding comments

Mathematical models have proved a useful tool for understanding the dynamics of virus infections [19] and have been widely applied to influenza infections [7] [20] [21] [22] [23] [24] [25] [26] [27] [28].

In this paper we focus on distinguishing between resource depletion and other mechanisms of control. We propose a means of rejecting resource depletion models. Namely we suggest that a substantial (>2 fold) increase in AUC with increase in I(0) is a criterion to reject resource depletion only models. Because of its simplicity and tractability, we focus on the Kermack McKendrick model. But we believe that our criterion is valid for many well mixed resource depletion models. In contrast a resource depletion model that is not well mixed may have different dynamics [29][30]. For example consider a spatial SIR model where there are islands of susceptibles and infection rarely jumps between islands. If higher I(0) corresponds to a larger number of infected islands at time zero, one might find a large increase in AUC with increasing I(0). But, while plausible for certain scenarios, we consider this model biologically implausible as a within host model for influenza.

It is important to note that our criterion for rejecting resource depletion does not work in the converse; near constant AUC with changing I(0) is not a strong argument in favor of resource depletion even when the resource depletion model fits pathogen load data well. In particular when there is no delay ($l_1+l_2=0$) our interferon model is mathematically equivalent to a target cell depletion model when it comes to fitting the time courses of infected and dead cells. But even in this case the interferon model allows interferon to play a much larger role than target cell depletion in terms of control of infection, so the models are conceptually quite different despite being in a sense mathematically equivalent. In this case distinguishing between the models would require additional data, different analyses or stronger assumptions.

Many factors such as resource depletion, innate and adaptive immunity can contribute to the control of infections and it is not always simple to determine their contributions. One approach to discriminating between models is to determine their ability to recapitulate the dynamics of the pathogen following a typical infection initiated by a fixed inoculum. However, the basic pattern of acute infections - an exponential growth phase followed by control and clearance - can be recapitulated by models with any of the above mechanisms. Consequently alternative approaches may be more powerful. One approach is to confront models with data showing how the pathogen dynamics change with innoculum dose [12]. Here we showed how such an analysis can allow rejection of the resource depletion hypothesis for influenza.

# Methods

## Experimental data

The experimental data comes from a previously published study. See [2] for details. The mouse experiments were approved by the Institutional Animal Care and Use Committee of Mount Sinai School of Medicine and conducted in accordance with their guidelines.

## Data analysis methods

To calculate AUCs from the Manicassamy et al data, replicates were averaged then the rectangle method was applied. Confidence intervals for the AUCs were calculated via bootstrap using 100000 resamples; bootstrap was stratified by dose group and time point. The interferon model was fit by minimizing

$\sum_{i}((ln(AUC_i) - ln(mAUC_i))^2) + \sum_{ij}((asinh(100 * I_{ij}) - asinh(100 * mI_{ij}))^2)$ where mAUC is the model AUC, I is the proportion of infected cells averaged across replicates, mI is the modeled proportion of infected cells, i represents the dose groups and j represents time points with data. This inverse hyperbolic sine transformation is similar to a linear transformation for small values of I and mI and similar to natural log transformation for values of I and mI that are larger than 0.01. For fitting the lowest dose group I=0 was imputed for days 5, 6 and 7.

# Acknowledgments

We thank Beth Kochin for insightful observations.

# Supporting information

S1. Appendix. Kermack McKendrick model can fit Saenz data.

S2. Appendix. Derivation of final size formula for the Kermack McKendrick model.

S3. Appendix. Final size of a virus explicit resource depletion model.

# S1. Kermack McKendrick model can fit Saenz data

Saenz et al [3] analyze primary influenza infection data from ponies. At days 2.5, 4.5 and 5.5 1.93%, 4.73% and 1.87% of lung epithelial cells were estimated to be infected. Saenz et al argue that this data is inconsistent with target cell depletion models because a majority of cells appear to have escaped infection. In fact this result can easily be explained by latent heterogeneity such that only a subset of cells are susceptible to infection. A Kermack McKendrick model with I(0)=5.99*10$^{-7}$, N=0.195, r=26.6 and b=1 fits the data almost perfectly; here the units are proportion of lung epithelial cells for I(0) and N and inverse days for r and b.

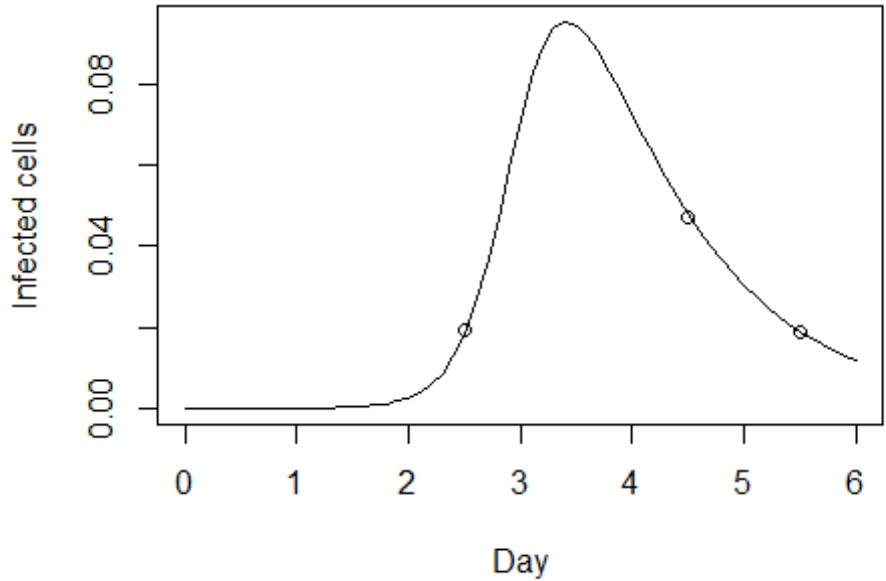

Saenz et al argue this data is inconsistent with target cell depletion being the only means of control. But in contradiction to this argument a Kermack McKendrick model fits the data almost perfectly.

## S2. Derivation of final size formula for the Kermack McKendrick model

$$dS/dt = -r * S * I$$

$$dI/dt = r * S * I - b * I$$

We can solve explicitly for I as a function of S

$$dI/dS = -1 + b/r/S$$

$$I = -S + b/r * ln(S) + constant$$

Solving for initial condition I(0)=N-S(0),

$$I = -S + b/r * ln(S) + N - b/r * ln(N - I(0))$$

At the end of infection I=0 and S=N-F where F is the final size so

$$0 = -N + F + b/r * ln(N - F) + N - b/r * ln(N - I(0))$$

$$0 = R_0 * F/N + ln(N - F) - ln(N - I(0))$$

since $R_0 = N*r/b$

Therefore

$$N - I(0) = (N - F) * e^{R_0*F/N}$$

$$-e^{-R_0} * R_0/N * (N - I(0)) = -e^{-R_0} * R_0/N * (N - F) * e^{R_0*F/N}$$

$$-e^{-R_0} * R_0 * (1 - I(0)/N) = R_0/N * (F - N) * e^{R_0*(F-N)/N}$$

$$W(-e^{-R_0} * R_0 * (1 - I(0)/N)) = R_0 * (F - N)/N$$

$$F = N * (1 + W(-e^{-R_0} * R_0 * (1 - I(0)/N))/R_0)$$

where W is the Lambert W function which has the property that $W(z*e^z)=z$. The Lambert W function has two branches since $z*e^z=z$ has up to two real solutions. The upper branch of the Lambert W function gives the correct value for F.

## S3. Final size of a virus explicit resource depletion model

A virus explicit model is shown below.

$$dV/dt = p * I - a * V$$

$$dS/dt = -r * V * S$$

$$dI/dt = r * V * S - b * I$$

$$dR/dt = b * I$$

Here, V is quantity of virus, S is the number of susceptible cells, I is the number of infected cells, and R is the number of dead cells.

The final size of this model is given by a modified Kermack McKendrick final size formula.

$$F = N * (1 + W(-R_0 * (1 - EIZ/N) * e^{-R_0})/R_0)$$

$$EIZ = I(0) + S(0) * (1 - e^{-r/a*V(0)})$$

Here, $N=S(0)+I(0)$, W is the upper branch of the Lambert W function, $R_0=N*p*r/a/b$ and EIZ can be interpreted as the dose of infected cells without virus that is equivalent to a dose of V(0) virus and I(0) infected cells. For V(0)=0 the above formula simplifies to the standard Kermack McKendrick final size formula.

The modified final size formula can be derived from a probabilistic argument. Let A be infectious contact with virus produced by infected cells, and let B be infectious contact with the initial stock of virus. Here infectious contact means contact that would have caused infection had the cell been susceptible.

$$N - F = S(0) * P(\neg A) * P(\neg B)$$

It can be shown that $P(\neg A) = e^{-R_0 * F/N}$ and $P(\neg B) = e^{-r/a * V(0)}$. Solving for F gives the modified final size formula. Notably this probabilistic argument and this modified final size formula apply also to more complicated resource depletion models; this includes models in which infected cells progress through multiple infectious stages and those in which certain cell populations are more infectious than others [15].